\documentclass[final,5p,times]{elsarticle}

\usepackage{amsmath,amssymb}
\usepackage{graphicx}
\usepackage{placeins}

\graphicspath{{./}{figures/}}
\usepackage{bm}
\usepackage{color}
\usepackage{comment}
\usepackage[hidelinks]{hyperref}

\journal{Journal of Magnetism and Magnetic Materials}
\biboptions{sort&compress}

\newcommand{\avg}[1]{\left\langle #1 \right\rangle}

\begin{document}

\begin{frontmatter}

\title{Learning-Performance Evaluation of a Physical Reservoir Based on a Vortex Spin-Torque Oscillator with a Modified Free Layer}

\author[ibaraki]{Kota Horizumi}
 \ead{24nd253n@vc.ibaraki.ac.jp}

\author[yamagata,tohoku]{Takahiro Chiba}

\author[ibaraki]{Takashi Komine\corref{cor1}}
\cortext[cor1]{Corresponding author.}

\address[ibaraki]{Graduate School of Science and Engineering, Ibaraki University, Hitachi, Ibaraki 316-8511, Japan}
\address[yamagata]{Department of Information Science and Technology, Graduate School of Science and Engineering, Yamagata University, Yonezawa, Yamagata 992-8510, Japan}
\address[tohoku]{Department of Applied Physics, Graduate School of Engineering, Tohoku University, Sendai, Miyagi 980-8579, Japan}

\begin{abstract}
In this study, we numerically evaluate the learning performance of a vortex spin-torque oscillator (VSTO) with a modified free layer, called a modified VSTO (m-VSTO), in which an additional layer (AL) of smaller radius is stacked on the free layer, for physical reservoir computing.
The vortex-core dynamics are computed using the Thiele equation incorporating the potential deformation induced by the AL.
We identify the edge of chaos from the maximal Lyapunov exponent and quantify the short-term memory capacity (STMC) as well as the information processing capacity (IPC) in a time-multiplexed reservoir scheme.
We find that the m-VSTO exhibits finite STMC and IPC in a low-current and low-field regime below the threshold current of the conventional VSTO, and can achieve up to approximately twice the IPC with about one quarter of the power consumption.
The pulse-width dependence of the IPC can be further interpreted by combining an analytical estimate of the transient time with Lyapunov-exponent data. Longer pulse widths promote stronger recovery-induced contraction toward the groove-trapped orbit over a wide range of subthreshold currents. In contrast, this difference in contraction rate becomes less pronounced near the threshold current, where the transient time increases rapidly.
Consequently, the IPC is enhanced in a stable regime with a negative Lyapunov exponent rather than exactly at the edge of chaos.
These results suggest that engineering the potential landscape and pulse-width-dependent recovery dynamics enables low-power spintronic physical reservoirs.
\end{abstract}


\end{frontmatter}

\section{Introduction}

Physical reservoir computing (PRC), which exploits nonlinear dynamics in physical systems for information processing, has attracted considerable attention as an energy-efficient hardware platform for machine learning.\cite{Lukosevicius2009, Tanaka2019}
As a widely discussed guideline, PRC performance is often reported to be high near the so-called edge of chaos (EoC), which refers to the transition region between ordered and chaotic dynamics.\cite{Langton1990, Boedecker2012}
To date, a variety of physical systems have been proposed as PRC candidates, including optical systems\cite{Wang2024} and mechanical systems.\cite{Sun2021}
Among them, spintronic devices are promising because they enable compact and high-speed information processing, and have therefore been actively investigated.\cite{Kanao2019,Akashi2020,Yamaguchi2023,Namiki2024,Tatsumi2025}
Although a direct comparison with the present study is difficult because the experimental configuration and the definition of the reservoir output differ\cite{Tsunegi2019,Ismail2019,Yamaguchi2020,Taniguchi2022,Shreya2023,Shen2024}, Tsunegi \textit{et al.}\cite{Tsunegi2019} experimentally demonstrated PRC using a vortex-type spin-torque oscillator, in which an electrical output signal containing information on the phase of the vortex-core gyration was employed, and showed that the system possesses learning capability.

However, because the conventional VSTO requires a finite threshold current to sustain self-oscillation, it is difficult to sufficiently reduce the operating current, which limits its energy efficiency.\cite{Pribiag2007,Yamamoto2016,Kamimaki2021}
Moreover, it has been pointed out that, in VSTOs, the short-term memory capacity (STMC), which quantifies how much past input a reservoir can retain,\cite{Inubushi2017}
and the information processing capacity (IPC), which evaluates how accurately the reservoir can realize complex nonlinear mappings,\cite{Dambre2012}
are not necessarily maximized at the EoC; thus, the relationship between nonlinearity and learning performance remains unclear.\cite{Teuscher2022,Imai2022}

In this paper, we systematically evaluate, through numerical simulations, the learning performance of the modified VSTO (m-VSTO) that we recently proposed.\cite{HorizumiJMSJ2025,HorizumiIEEE2025}
In an m-VSTO, the potential landscape is modified by the presence of a small-radius additional layer (AL) on the free layer, which endows the device with strong nonlinearity that can produce chaotic dynamics under an applied ac magnetic field, while also enabling self-oscillation at lower currents than in a conventional VSTO.
(i) We first compute the dependence of the maximal Lyapunov exponent,\cite{StrogatzBook2001}
a standard indicator of chaos, on the dc current and external magnetic field to identify the EoC region exhibited by the m-VSTO.
(ii) We then evaluate the STMC and IPC in the same parameter space and examine in detail the relationship between the Lyapunov exponent and these learning-performance metrics.
(iii) To explain why high STMC/IPC can emerge away from the EoC,
we analyze the recovery dynamics of the m-VSTO after input perturbations by combining an analytical transient-time estimate with the pulse-width dependence of the Lyapunov exponent.
Based on this relation, we clarify how pulse-width-dependent contraction toward the groove-trapped orbit supports enhanced learning performance under low-current operation.

\section{Model and numerical methods}

\begin{figure}[!tbp]
  \centering
  \includegraphics[width=0.92\columnwidth,height=0.40\textheight,keepaspectratio]{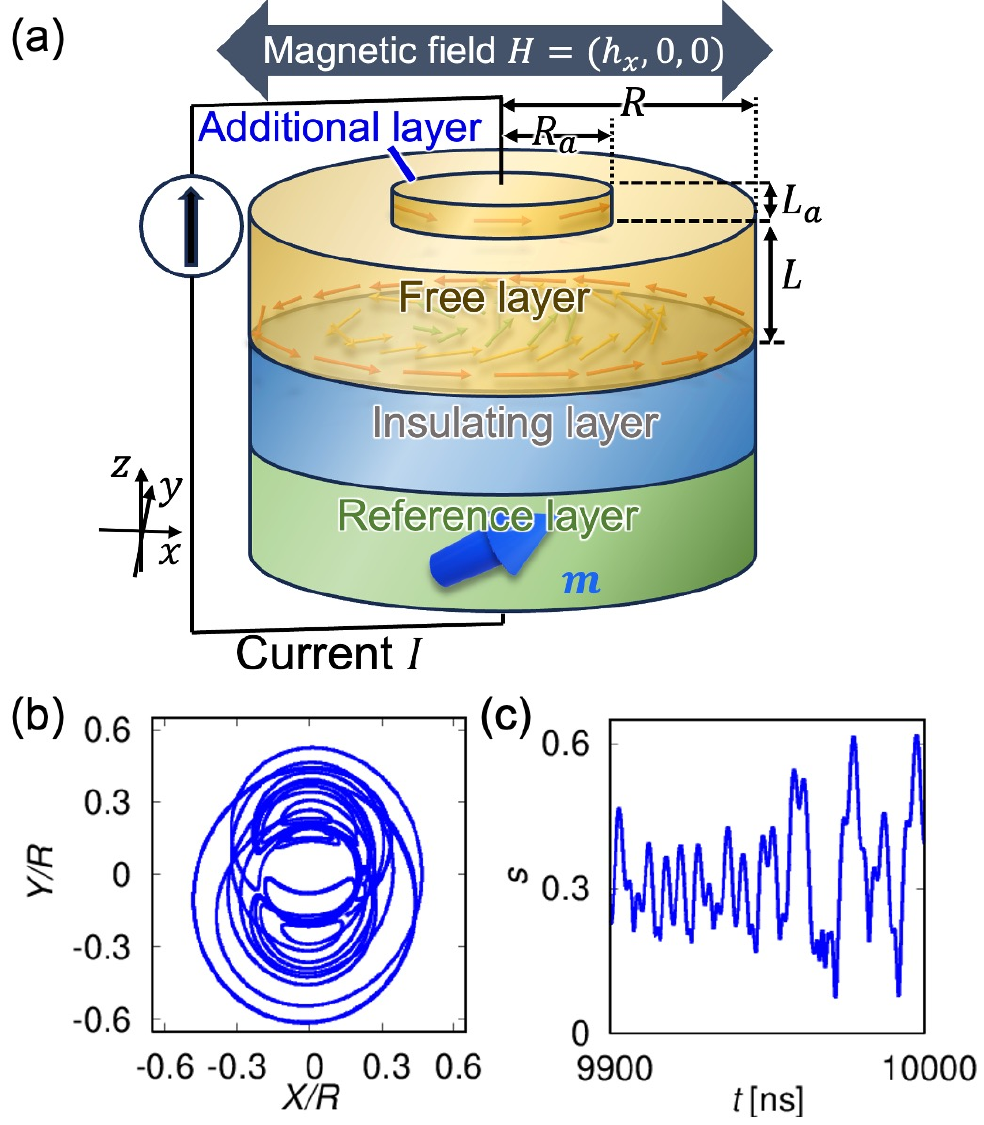}
  \caption{(a) Schematic of the modified vortex spin-torque oscillator (m-VSTO).
A circular ferromagnetic additional layer (AL) of radius $R_a$ and thickness $L_a$ is concentrically stacked on the free layer (radius $R$, thickness $L$) of a ferromagnet/insulator/ferromagnet trilayer.
A dc current $I$ is injected perpendicular to the film plane, and an in-plane magnetic field $\bm{H}=(h_x,0,0)$ is applied along the $x$ axis; $\bm{m}$ denotes the magnetization direction of the reference layer.
(b),(c) Representative vortex-core dynamics under an $x$-directed sinusoidal ac magnetic field at $I=0$:
(b) phase portrait of the normalized vortex-core position $(X/R,\,Y/R)$ after discarding an initial transient and
(c) the corresponding time trace of the normalized gyration amplitude $s(t)$.
The aperiodic trajectory and irregular amplitude fluctuations indicate a chaotic response enabled by the AL-induced deformation of the confinement potential.}
  \label{fig:fig1}
\end{figure}

As shown in Fig.~\ref{fig:fig1}(a), we consider an m-VSTO in which a circular ferromagnetic AL made of the same material as the free layer, but with a smaller radius than the free layer, is concentrically stacked on the free layer with the radius $R$ and the thickness $L$ of a VSTO composed of a ferromagnet/insulator/ferromagnet trilayer.
We employ a Cartesian coordinate system whose origin is at the center of the free layer and denote the basis vectors by $\boldsymbol{e}_k\ (k=x,y,z)$.
In the presence of a dc-current-induced spin torque and an external magnetic field $\boldsymbol{H}$, the motion of the vortex-core position $\boldsymbol{X}$ is described by the Thiele equation:\cite{Thiele1973,HorizumiJMSJ2025,HorizumiIEEE2025,Imai2022,Dussaux2012,Grimaldi2014,Tsunegi2021}
\begin{flalign}
&-G\boldsymbol{e}_z\times\dot{\boldsymbol{X}}
-|D|(1+\xi s^2)\dot{\boldsymbol{X}}
-\frac{\partial W(s)}{\partial \boldsymbol{X}}\notag\\
&
+a_J J m_z\,\boldsymbol{e}_z\times \boldsymbol{X}+c a_J R_0 m_x\,\boldsymbol{e}_x
+c\mu^{*}\,\boldsymbol{e}_z\times\boldsymbol{H}
=\boldsymbol{0}\,,
\label{Thiele}
\end{flalign}
where $\boldsymbol{X}=(X,Y,0)$ is the vortex core position vector, $\dot{\boldsymbol{X}}$ is the velocity vector, and $s=\sqrt{X^2+Y^2}$ is the distance of the vortex core from the center of the free layer.
Here, $W(s)$ is the potential function and $J=I/(\pi R^2)$ is the current density.
The coefficient $a_J=\pi\hbar \sigma/(2e)$ characterizes the spin-transfer torque, and the other coefficients are given by
$G=\frac{2\pi pML}{\gamma}$,
$D=-\frac{2\pi \alpha ML}{\gamma}\left(1-\frac{1}{2}\ln\frac{R_0}{R}\right)$, and
$\mu^{*}=\pi MLR$.
Here, $M$ is the saturation magnetization, $\gamma$ is the gyromagnetic ratio, $\alpha$ is the Gilbert damping constant, $R_0$ is the vortex-core radius, $\xi$ is the nonlinear damping coefficient, $\sigma$ is the spin polarization, $\boldsymbol{m}=(m_x,m_y,m_z)$ is the magnetization direction of the reference layer, $p$ is the polarity, and $c$ is the chirality.
The external magnetic field is taken as $\boldsymbol{H}=(h_x,0,0)$.
The VSTO consists of FeB/MgO/CoFeB multilayers.
Unless otherwise stated, the parameters in Eq.~(\ref{Thiele}) are set to
$M=1300~\mathrm{emu/cm^3}$,
$\gamma=1.764\times10^{7}~\mathrm{rad/(Oe\,s)}$,
$\alpha=0.01$, $\xi=2$, and the spin polarization $P=0.7$.
We set the vortex polarity and chirality to $p=1$ and $c=1$,
and use the vortex-core radius $R_0=10~\mathrm{nm}$.
The free-layer radius and thickness are $R=187.5~\mathrm{nm}$ and $L=5~\mathrm{nm}$.
For the m-VSTO, an additional FeB layer is concentrically stacked on the free layer
with radius $R_a=40~\mathrm{nm}$ and thickness $L_a=L/3$.
The reference-layer magnetization direction is taken as
$\bm{m}=(m_x,0,m_z)=(\sin(\pi/3),0,\cos(\pi/3))$.
These parameter values follow the established VSTO/m-VSTO modeling in the literature.
\cite{HorizumiIEEE2025,HorizumiJMSJ2025,Imai2022}
Possible changes in the effective Thiele coefficients such as $G$ and $D$ due to the AL are neglected here because they are not expected to change the qualitative role of the AL-induced deformation of $W(s)$.

For a conventional VSTO,\cite{Imai2022} the potential can be approximated by a single-well form with a minimum at the center, $W(s)\simeq \frac{\kappa}{2}s^2+\frac{\kappa'}{4}s^4$.
In contrast, the potential of the m-VSTO takes a Mexican-hat-like (wine-bottle-like) form, featuring an annular minimum near the edge of the additional layer.
The modified potential profile $W(s)$ for the vortex core was obtained from micromagnetic simulations\cite{MuMax3}
using the two-vortices ansatz,\cite{Guslienko2014}
assuming that the magnetizations in the free layer and the AL are strongly coupled via exchange interaction.
The effect of the modified potential was incorporated into the Thiele equation through the restoring force term $-\partial W(s)/\partial \bm{X}$.

We have systematically investigated how the AL deforms the confinement potential and thereby endows the m-VSTO with novel functionalities.\cite{HorizumiJMSJ2025,HorizumiIEEE2025}
In conventional VSTOs, inducing chaotic dynamics typically requires engineered driving schemes, such as nanocontact geometries,\cite{PetitWatelot2012NatPhys,Devolder2019PRL}
delayed feedback,\cite{Kamimaki2021}
or random magnetic fields.\cite{Imai2022}
By contrast, in the m-VSTO the AL reshapes the vortex-core confinement potential, effectively enhancing the nonlinearity of the restoring force and enabling chaotic vortex-core motion even without a dc current.
As a representative example, we apply an $x$-directed sinusoidal field
$h_x(t)=h_0\sin(2\pi f t)$ with $f=100~\mathrm{MHz}$ and $h_0=25~\mathrm{Oe}$ at $I=0$,
for which the trajectory becomes aperiodic and the gyration amplitude exhibits irregular fluctuations, indicating a chaotic response [Fig.~\ref{fig:fig1}(b),(c)].
This sinusoidal-field example is used to illustrate the strong nonlinearity induced by the AL, whereas the reservoir-performance evaluations below are carried out using random-pulse inputs.
Furthermore, when the initial vortex-core position is chosen near the annular groove (potential minimum) at $s\simeq s_0$,
sustained gyration emerges even below the threshold current of the conventional VSTO, $I_{\mathrm{th}}$ ($I<I_{\mathrm{th}}$),
and the below-threshold oscillation frequency $f_{\mathrm{below}}(I)$ depends approximately linearly on the applied current $I$.
Unless otherwise stated, all m-VSTO simulations presented below are initialized with the vortex core placed near the annular minimum, \textit{i.e.}, $s(0)\simeq s_0$, so that the dynamics starts from the groove-trapped state.
As $I\to I_{\mathrm{th}}$, $f_{\mathrm{below}}(I)$ continuously connects to the conventional oscillation frequency $f_{\mathrm{osci}}$,
and for $I>I_{\mathrm{th}}$ the oscillation frequencies of the conventional VSTO and the m-VSTO coincide.
In practice, we prepare the groove-trapped initial state by a short preconditioning step:
a dc-current drive is first applied to excite vortex gyration, and the core subsequently relaxes into the annular minimum, providing a reproducible initial condition.\cite{HorizumiIEEE2025}
These characteristics---self-sustained dynamics in the low-current regime and a strong nonlinear response absent in conventional VSTOs---are well suited to PRC, where nonlinearity and transient responses serve as computational resources; accordingly, in this study we quantitatively evaluate how these properties influence learning-performance metrics.

The Thiele equation was numerically solved to obtain the vortex-core dynamics, and the maximal Lyapunov exponent was computed using the Shimada--Nagashima method\cite{ShimadaNagashima1979} to identify the EoC.  This Lyapunov-exponent-based classification has also been used to distinguish input-driven synchronization and chaos in spintronic magnetization dynamics.\cite{TaniguchiJMMM2022}  In brief, for a reference trajectory and a nearby perturbed trajectory whose separation after the $i$th renormalization interval is $D_i$, the maximal Lyapunov exponent is evaluated as
\begin{equation}
\lambda = \lim_{N\to\infty}\frac{1}{N\Delta t}\sum_{i=1}^{N}\ln\left(\frac{D_i}{\epsilon}\right),
\label{eq:lyapunov_main}
\end{equation}
where $\epsilon$ is the prescribed perturbation distance and $\Delta t$ is the renormalization interval.  The detailed renormalization procedure is summarized in \ref{app:lyapunov}.
Furthermore, to evaluate the learning performance, we used a fixed-pulse-width external magnetic field as the input, the normalized radius distance $s$ as the output, and adopted a time-division multiplexing scheme.
The output within each pulse interval was divided into 50 segments, simulating 50 virtual nodes.
The short-term memory capacity quantifies the ability of the reservoir to reconstruct past inputs, whereas the information processing capacity evaluates its capability to perform nonlinear computational tasks. The detailed definitions of the STMC and IPC, together with the target functions used in this work, are given in \ref{app:learning_performance}. 
For the evaluations of STMC and IPC, the input sequences for training and testing were generated independently; after training the readout, we re-run the device dynamics from the same prescribed initial condition to construct the test reservoir states, so that no explicit washout is required in our protocol.
The readout weights were obtained by ordinary least squares using the Moore-Penrose pseudoinverse.
Unless otherwise stated, the re-initialization places the vortex core near the annular groove, $s(0)\simeq s_0$, to start from the groove-trapped state.
The hyperparameters used in the present calculations were chosen as representative values based on previous VSTO reservoir studies and computational feasibility, and were not globally optimized.

\section{Lyapunov exponent and edge of chaos}

\begin{figure}[!b]
  \centering
  \includegraphics[width=0.98\columnwidth,height=0.28\textheight,keepaspectratio]{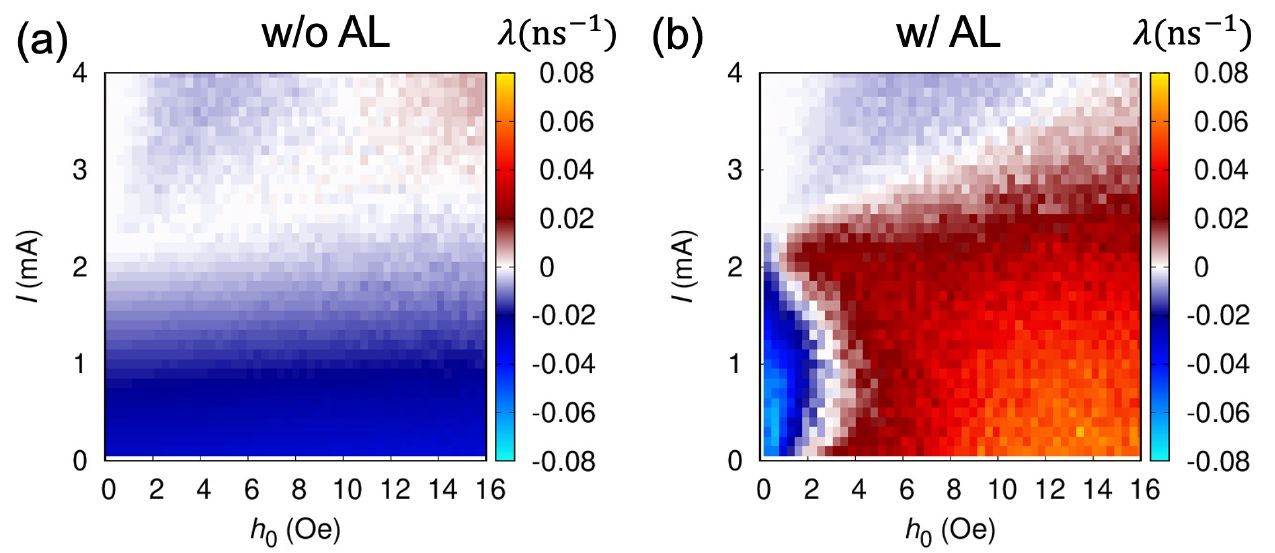}
\caption{Dependence of Lyapunov exponent $\lambda$ on input strength $h_0$ (Oe) and current $I$ (mA) for (a) without AL and (b) with AL radius $R_a=40$\,nm.
White indicates a region where $\lambda$ is very close to zero, which we regard as corresponding to the edge of chaos.
}
  \label{fig:fig2}
\end{figure}

We first compute the dependence of the maximal Lyapunov exponent on the dc current and external magnetic field to identify the EoC region exhibited by the m-VSTO.
Figure~\ref{fig:fig2} shows the dependence of the Lyapunov exponent $\lambda$ on the external magnetic field amplitude and the applied dc current for a conventional VSTO and an m-VSTO.
In these calculations, the $x$ component of the external magnetic field $\boldsymbol{H}$ is given by $h_x(t)=h_0 r(t)$, where $r(t)$ is a random pulse train with a fixed pulse width of $3~\mathrm{ns}$ generated from a uniform distribution over $[-1,1]$.
The horizontal axis denotes the field amplitude $h_0$ ranging from 0 to 16~Oe, and the vertical axis denotes the applied current $I$ ranging from 0 to 4~mA.
A positive $\lambda$ indicates chaotic dynamics, whereas a negative $\lambda$ indicates nonchaotic (periodic) dynamics.
We color-code the regions with $\lambda>0$ in red and those with $\lambda<0$ in blue, and we mark regions where $\lambda$ is very close to zero $|\lambda|\simeq0$, in white.
In the present calculations, the Lyapunov exponent was evaluated over a finite calculation time of $T_{\mathrm{cal}}=100~\mu\mathrm{s}=1.0\times10^{5}~\mathrm{ns}$. Because the Lyapunov exponent has the dimension of inverse time, the numerical resolution associated with this calculation time is estimated as
$\delta\lambda=\frac{1}{T_{\mathrm{cal}}}\simeq1.0\times10^{-5}~\mathrm{ns}^{-1}$.
Accordingly, 
$
|\lambda|\leq\delta\lambda\simeq1.0\times10^{-5}~\mathrm{ns}^{-1}
$
is shown in white and is operationally defined as the edge of chaos in this work.

For the conventional VSTO as shown in Fig.~\ref{fig:fig2}(a), as reported previously, self-oscillation emerges above the threshold current $I_{\mathrm{th}}\simeq 2.1~\mathrm{mA}$, and chaos is induced mainly at large field amplitudes.
In contrast, for the m-VSTO as shown in Fig.~\ref{fig:fig2}(b), the EoC already appears at currents below the threshold of the conventional device, and chaos is induced at lower field amplitudes than in the conventional VSTO.
As a result, the m-VSTO exhibits a broad EoC region in the low-current and low-field regime.

\section{Learning performance}

\begin{figure}[!tbp]
  \centering
  \includegraphics[width=0.98\columnwidth,height=0.40\textheight,keepaspectratio]{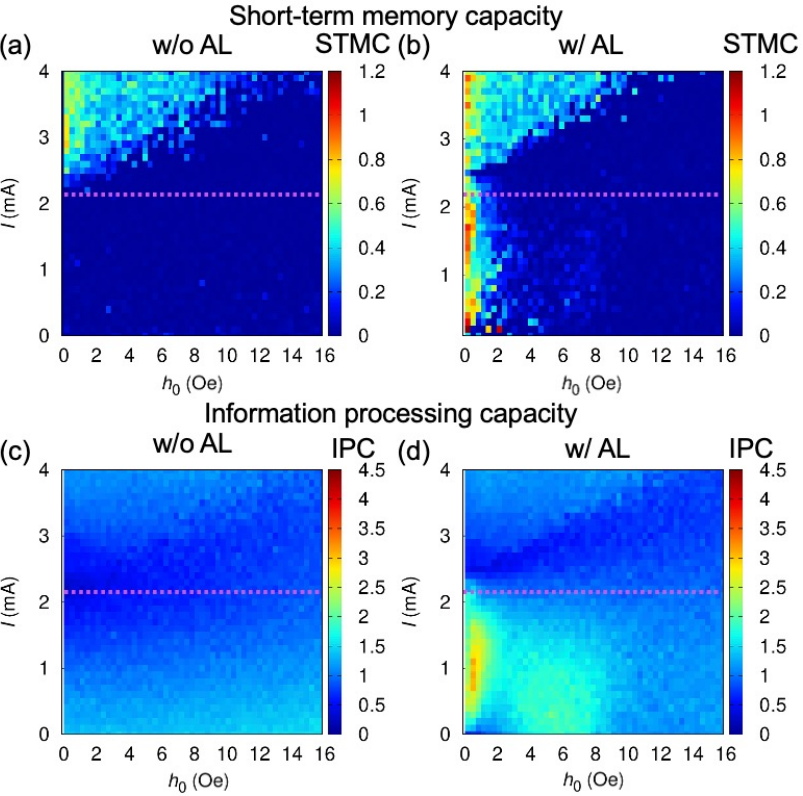}
  \caption{Dependence of short-term memory capacity (STMC) and information processing capacity (IPC) on input strength $h_0$ (Oe) and current $I$ (mA) for (a)(c) without AL and (b)(d) with AL radius $R_a=40$\,nm. The purple dashed line indicates the threshold current of a conventional VSTO.}
  \label{fig:fig3}
\end{figure}

Next, we evaluate the STMC and IPC in the same parameters and examine in detail the relationship between the Lyapunov exponent and these learning-performance metrics.
Figure~\ref{fig:fig3} shows the dependence of the learning-performance metrics, STMC and IPC, on the external-field strength and the dc current.
In all calculations, the pulse width was fixed at $3~\mathrm{ns}$, whereas the input sequences used for the STMC and IPC evaluations were different.
For the STMC calculation, we set the sequence length to $N=1000$, where $N$ is the number of input signals, and defined the external field as $\boldsymbol{H}(t)=(h_0 r_b(t),0,0)$, where $r_b(t)$ is a binary random function that takes values 0 or 1.
For the IPC calculation, we increased the sequence length to $N=5000$ so that the estimated IPC value sufficiently converges, and we defined $\boldsymbol{H}(t)=(h_0 r(t),0,0)$ using a random function $r(t)$ drawn from a uniform distribution on $[-1,1]$ to preserve orthogonality among nonlinear functions.
Since both STMC and IPC depend on the evaluation protocol, their absolute values are not universal performance measures. However, when evaluated using an identical protocol, they provide quantitative metrics for comparing the computational capability of different reservoir systems. Accordingly, the distributions shown in Fig. 3 should be interpreted as a direct comparison of the computational performance resulting from the different device geometries.

Comparing Figs.~\ref{fig:fig3}(a) and \ref{fig:fig3}(b), which show the results for the STMC, we find that the capacity is almost zero below the threshold current for the conventional VSTO, whereas the m-VSTO exhibits a finite value under the same conditions.
Above the threshold current, the influence of the AL becomes small, and the results for the two devices nearly coincide.
Similar trends are observed for the information processing capacity in Figs.~\ref{fig:fig3}(c) and \ref{fig:fig3}(d).
In particular, the IPC of the m-VSTO is maximized around $h_0\simeq 1~\mathrm{Oe}$ and $I\simeq 1~\mathrm{mA}$, reaching approximately twice the value of the conventional VSTO while consuming about one quarter of the power.
Here the power consumption is estimated from Joule heating, $P\simeq I^2R$, neglecting the bias dependence of the device resistance and assuming comparable effective resistance values for the conventional VSTO and the m-VSTO. Therefore, operating the m-VSTO at $I\simeq 1~\mathrm{mA}$, which is about half of the threshold current of the conventional VSTO ($I_{\mathrm{th}}\simeq 2~\mathrm{mA}$), corresponds to approximately one quarter of the power consumption. This estimate should be regarded as an order-of-magnitude comparison, because possible changes in the effective resistance due to the AL are neglected.

A comparison between Fig.~\ref{fig:fig2} and Figs.~\ref{fig:fig3}(a) and \ref{fig:fig3}(b) indicates that ${\mathrm{STMC}}$ decreases as the system approaches the EoC and becomes almost zero in the chaotic regime.
By contrast, ${\mathrm{STMC}}$ tends to increase in regions where $\lambda$ is negative with a large magnitude.
The same tendency is also observed for the IPC, suggesting that the learning performance is enhanced not near the EoC but rather in the regime where the Lyapunov exponent takes the most negative values.
This trend suggests that, in our pulsed-drive and time-multiplexed setting, the performance is governed not by proximity to the EoC per se but by how the pulse width matches the transient time of the below-threshold dynamics, which we discuss next.

\section{Transient-time interpretation and pulse-width dependence}

\begin{figure}[!tbp]
  \centering
  \includegraphics[
    width=0.98\columnwidth,
    height=0.44\textheight,
    keepaspectratio
  ]{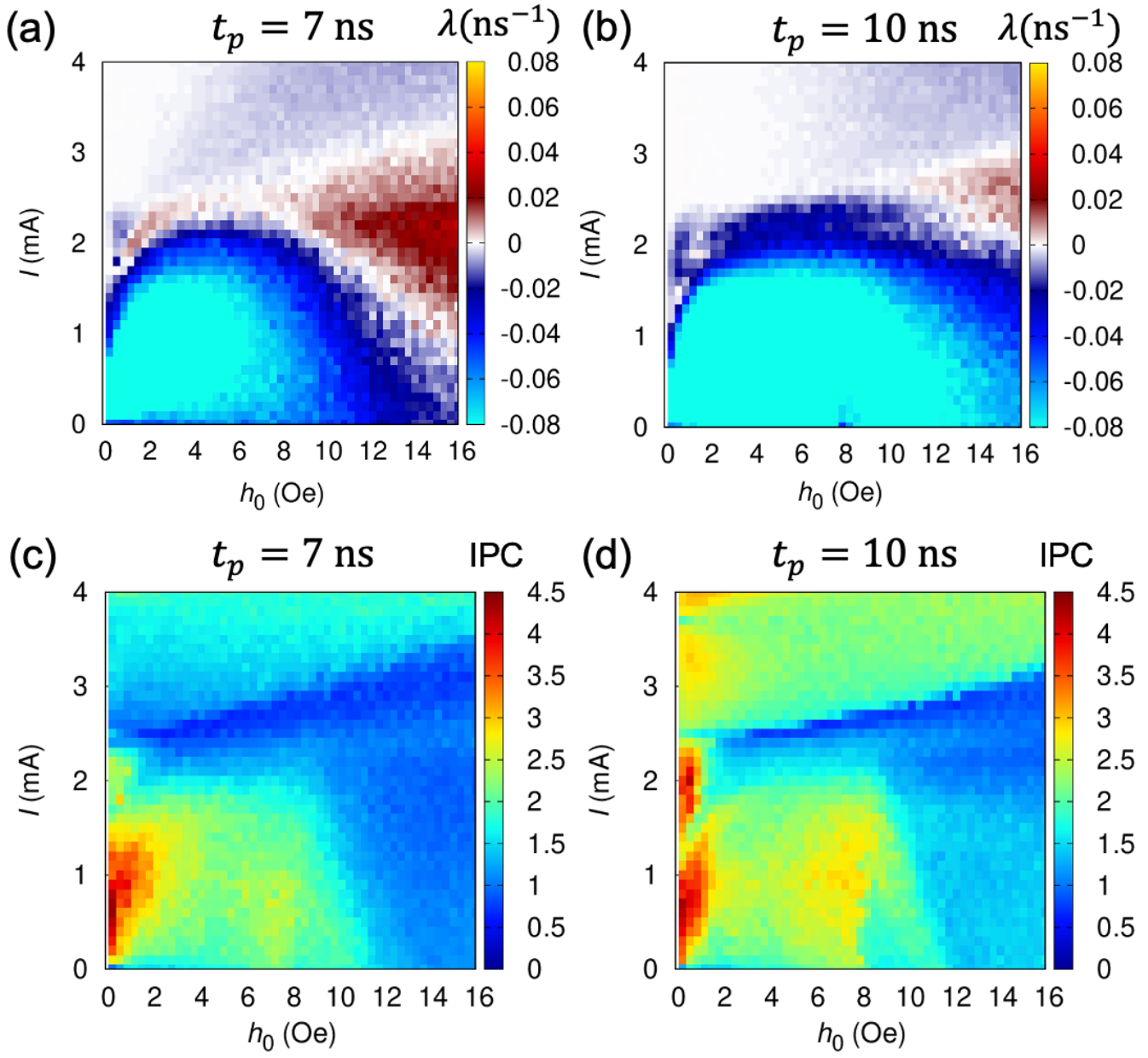}
  \caption{Pulse-width dependence of the Lyapunov exponent and IPC for the m-VSTO with $R_a=40$\,nm. (a),(b) Lyapunov exponent maps for $t_p=7$ and $10~\mathrm{ns}$, respectively. (c),(d) Corresponding IPC maps for $t_p=7$ and $10~\mathrm{ns}$. The comparison shows that a longer pulse width broadens the stable region with $\lambda<0$ below the threshold current and expands the high-IPC region in the same low-current regime.}
  \label{fig:fig4}
\end{figure}

To clarify why high STMC/IPC appears in a regime away from the EoC, we discuss how the interplay between the below-threshold recovery time and the input pulse width determines the sign and magnitude of the Lyapunov exponent, and we provide a guideline for choosing pulse widths that enhance learning performance under low-current operation.
In our time-multiplexed scheme, a sufficiently negative $\lambda$ suppresses divergence of nearby trajectories under pulsed perturbations, while a finite recovery time allows the reservoir state to retain information over multiple input steps; thus, the key dynamical factor is the recovery process within each input interval, rather than exact operation at the EoC.
Because the Lyapunov exponent quantifies whether the separation between two nearby trajectories increases or decreases exponentially, it reflects how strongly perturbations are contracted during the pulse interval and therefore characterizes the effective recovery of the dynamics toward the groove-trapped orbit.
Figure~\ref{fig:fig4} provides a two-dimensional parameter-space view of the pulse-width dependence of the stability and learning performance using representative cases. In Figs.~\ref{fig:fig4}(a) and \ref{fig:fig4}(b), the Lyapunov exponent maps are shown for $t_p=7$ and $10~\mathrm{ns}$, respectively, while Figs.~\ref{fig:fig4}(c) and \ref{fig:fig4}(d) show the corresponding IPC maps. These maps demonstrate that increasing the pulse width broadens the stable region with $\lambda<0$ in the subthreshold regime and expands the high-IPC region, rather than simply increasing the peak IPC value.
To connect this trend to a characteristic relaxation time, we linearize the Thiele equation near the annular groove formed by the additional layer and estimate the transient time in the below-threshold regime, \textit{i.e.}, $I<I_{\rm th}$.
Using the polar representation $X+iY=se^{i\psi}$, we approximate the potential near the groove position $s_0$ as $W(s)\simeq (\kappa_0/2)(s-s_0)^2$ and treat the damping locally as a constant,
$D_0=|D|(1+\xi s_0^2)$, while defining $A_J(I)=a_Jp_zJ$.
Under these approximations, Eq.~(\ref{Thiele}) can be rewritten as
\begin{equation}
\left\{
\begin{aligned}
& G s\,\dot{\psi} - D_0\,\dot{s} - \kappa_0\,(s - s_0) = 0, \\
&-\,G\,\dot{s} - D_0\,s\,\dot{\psi} + A_J(I)\,s = 0.
\end{aligned}
\right.
\end{equation}
Solving these equations for $\dot{s}$ yields
\begin{equation}
    \dot{s}=\frac{(A_J(I)G- D_0\kappa_0)s+ D_0\kappa_0s_0}{G^2+ D_0^2}\,.
    \label{sdot}
\end{equation}
The steady-state radius $s^*$ satisfying $\dot{s}=0$ is then
\begin{equation}
    s^* = \frac{D_0\kappa_0s_0}{D_0\kappa_0-A_J(I)G}\,.
    \label{stady}
\end{equation}
Linearizing around the steady orbit $s=s^*$ by setting $s=s^*+\delta s$, we obtain
\begin{equation}
    \dot{\delta s}=\frac{A_J(I)G-D_0\kappa_0}{G^2+D_0^2}\delta s\,.
\end{equation}
The perturbation therefore decays as $\delta s(t)=\delta s(0)e^{-\gamma t}$ when $D_0\kappa_0>A_J(I)G$, and the transient time is given by $\tau_{\rm below}=1/\gamma$, i.e.,
\begin{equation}
\tau_{\rm below}(I)\simeq \frac{G^2+D_0^2}{D_0\kappa_0-A_J(I)G}\,.
\end{equation}
Furthermore, by expressing the below-threshold gyration frequency as $f_{\rm below}(I)=A_J(I)/(2\pi D_0)$ and using the fact that $\kappa_0/(2\pi G)$ corresponds to the conventional VSTO oscillation frequency $f_{\mathrm{conv}}$ in the above-threshold regime, we can rearrange the result into the compact form
\begin{equation}
\tau_{\mathrm{below}}(I)=\tau_0\left(1-\frac{f_{\mathrm{below}}(I)}{f_{\mathrm{conv}}}\right)^{-1}\,,
\end{equation}
where $\tau_0=(G^2+D_0^2)/(D_0\kappa_0)$ is the transient time at zero current within this linearized model.
This estimate is consistent with Fig.~\ref{fig:fig4}: it explains why the $\lambda<0$ region expands for longer pulse widths and why the high-IPC region is broadened in the same subthreshold parameter range.

\begin{figure}[!tbp]
  \centering
  \includegraphics[width=0.98\columnwidth,height=0.30\textheight,keepaspectratio]{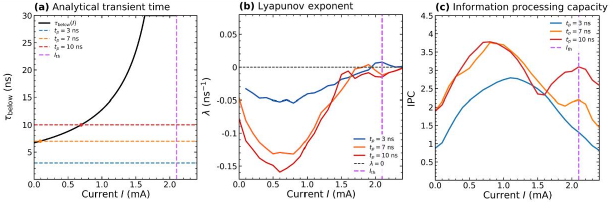}
  \caption{Current dependence of the analytical transient time, Lyapunov exponent, and IPC at a representative input amplitude $h_0=0.96~\mathrm{Oe}$. (a) Analytical transient time $\tau_{\rm below}(I)$ estimated from the linearized subthreshold dynamics. (b) Maximal Lyapunov exponent obtained from numerical simulations. (c) IPC obtained from the same data set. The vertical purple dashed line indicates the threshold current $I_{\rm th}$ of the conventional VSTO. For $t_p=7$ and $10~\mathrm{ns}$, the IPC remains high over a broader subthreshold current interval than for $t_p=3~\mathrm{ns}$, while the dynamics remains mostly stable with $\lambda<0$.}
  \label{fig:fig5}
\end{figure}

After this two-dimensional overview, Fig.~\ref{fig:fig5} extracts a representative low-field slice at $h_0=0.96~\mathrm{Oe}$ and compares the analytical transient time, the maximal Lyapunov exponent, and the IPC as functions of the dc current.
Figure~\ref{fig:fig5}(a) shows that the analytical transient time $\tau_{\rm below}(I)$ rapidly increases as the current approaches the threshold current.
This increase means that the recovery toward the groove-trapped orbit becomes slower near the threshold.
Figure~\ref{fig:fig5}(b) shows that, in a broad subthreshold current range, the magnitude of the negative Lyapunov exponent tends to be larger for longer pulse widths.
This behavior is interpreted as a larger recovery fraction within one input interval: for a short pulse width such as $t_p=3~\mathrm{ns}$, the next input is applied before the vortex-core orbit has sufficiently recovered from the displacement caused by the previous input, whereas for a longer pulse width such as $t_p=10~\mathrm{ns}$, the orbit has more time to contract back toward the groove-trapped orbit.
Near the threshold current, however, $\tau_{\rm below}(I)$ becomes so long that the recovery within a finite pulse interval is incomplete for all pulse widths, and the differences among the Lyapunov-exponent curves become less pronounced.
Consistently, Fig.~\ref{fig:fig5}(c) shows that the IPC is enhanced in the stable subthreshold region with $\lambda<0$, especially for $t_p=7$ and $10~\mathrm{ns}$, where the high-IPC region is broader than that for $t_p=3~\mathrm{ns}$.
These results indicate that the enhanced IPC below threshold is not determined by proximity to the edge of chaos alone or by the peak value of IPC alone.
Rather, it is supported by an appropriate balance between stable contraction toward the groove-trapped orbit and a recovery time that is long enough to preserve transient memory but short enough to allow contraction within the input interval.
Therefore, by appropriately tuning the input pulse width, the m-VSTO can operate as a physical reservoir with high learning performance in the low-current and low-field regime.

\FloatBarrier

\section{Conclusion}

In summary, we systematically evaluated the vortex-core dynamics and the learning performance of an m-VSTO as a physical reservoir by numerical simulations based on the Thiele equation.
The AL on the free layer transforms the potential landscape into a Mexican-hat-like form, so that oscillations arise even in the below-threshold, low-current and low-field regime where a conventional VSTO exhibits little dynamics, and chaotic behavior is also observed.
In this low-current and low-field regime, both STMC and IPC take finite values, and, in particular, around $h_0\simeq 1~\mathrm{Oe}$ and $I\simeq 1~\mathrm{mA}$, we found that the m-VSTO achieves an IPC approximately twice that of a conventional VSTO with about one quarter of the power consumption.
We further showed that the pulse-width dependence of IPC can be understood from the recovery dynamics after input perturbations: longer pulse widths lead to stronger contraction toward the groove-trapped orbit over a broad subthreshold current range, as reflected in more negative Lyapunov exponents, whereas this pulse-width dependence weakens near the threshold because the analytical transient time rapidly increases.
The enhanced IPC is therefore obtained in a stable regime with $\lambda<0$ and is associated with recovery-induced contraction and transient memory, rather than with exact operation at the edge of chaos.
These results indicate that the coordinated design of the potential landscape and the input pulse width is key to improving the learning performance and reducing the power consumption of m-VSTO reservoirs.
Optimization of the AL geometry and input scheme, including possible current- or voltage-driven operation, remains an important subject for future work.
Overall, our study demonstrates the feasibility of energy-efficient physical reservoirs based on vortex spin-torque oscillators.

\section*{Declaration of competing interest}
The authors declare that they have no known competing financial interests or personal relationships that could have appeared to influence the work reported in this paper.

\section*{Data availability}
Data will be made available on reasonable request.

\section*{Acknowledgments}
This work was partly supported by Grants-in-Aid for Scientific Research (Grants No. 24K00916, 24H00018, 22K14591, 24K06912) from the Japan Society for the Promotion of Science and Amano Institute of Technology.
The authors thank Dr. Tomohiro Taniguchi (AIST) for helpful advice and discussions on the STMC calculation method.
This work was partially performed under the Research Program of ``Dynamic Alliance for Open Innovation Bridging Human, Environment and Materials'' in ``Network Joint Research Center for Materials and Devices.''

\appendix

\section{Evaluation method of the Lyapunov exponent}
\label{app:lyapunov}

The maximal Lyapunov exponent was evaluated by the Shimada--Nagashima renormalization method.  Let $\bm{x}(t)$ denote the state vector of the vortex-core dynamics and let $\bm{x}^{(1)}(t)$ be a slightly perturbed state satisfying
\begin{equation}
\left|\bm{x}^{(1)}(t_0)-\bm{x}(t_0)\right|=\epsilon,
\end{equation}
where $\epsilon$ is a sufficiently small fixed distance in phase space.  After evolving both trajectories for a time step $\Delta t$, the distance between them becomes
\begin{equation}
D_1=\left|\bm{x}^{(1)}(t_0+\Delta t)-\bm{x}(t_0+\Delta t)\right|.
\end{equation}
The instantaneous expansion factor is then $D_1/\epsilon$.  The perturbed trajectory is subsequently renormalized to keep the distance from the reference trajectory equal to $\epsilon$,
\begin{equation}
\bm{x}^{(2)}(t_0+\Delta t)
=
\bm{x}(t_0+\Delta t)
+
\epsilon
\frac{\bm{x}^{(1)}(t_0+\Delta t)-\bm{x}(t_0+\Delta t)}{D_1}.
\end{equation}
Repeating this evolution and renormalization procedure, we obtain a sequence of distances $D_i$.  The maximal Lyapunov exponent is calculated as
\begin{equation}
\lambda
=
\lim_{N\to\infty}
\frac{1}{N\Delta t}
\sum_{i=1}^{N}
\ln\left(\frac{D_i}{\epsilon}\right).
\label{eq:app_lyapunov}
\end{equation}
In the present simulations, the time step used for the Lyapunov-exponent evaluation was chosen to be sufficiently smaller than the input pulse width and the characteristic timescale of the vortex-core motion.  A positive value of $\lambda$ indicates sensitive dependence on initial conditions, whereas a negative value indicates stable contraction of nearby trajectories under the same input sequence.

\section{Evaluation method of learning-performance metrics}
\label{app:learning_performance}

We describe the calculation of the STMC and IPC used in the main text.  The input sequence is denoted by $\{q_i\}_{i=1}^{N}$, where $N$ is the number of input pulses.  The output signal of the m-VSTO within each input-pulse interval is divided into $N_{\mathrm{node}}$ segments, and the resulting time-multiplexed reservoir state at the $i$th input step is written as
\begin{equation}
\bm{u}_i=(u_i^1,u_i^2,\ldots,u_i^{N_{\mathrm{node}}}).
\end{equation}
Collecting the reservoir states for all input steps gives the state matrix
\begin{equation}
U=(\bm{u}_1,\bm{u}_2,\ldots,\bm{u}_N)^{\mathrm{T}}.
\end{equation}
In the numerical implementation, a bias column is appended to $U$.  For a given target sequence $\bm{T}$, the linear readout weight vector $\bm{w}$ is determined by ordinary least squares,
\begin{equation}
\bm{w}=
\operatorname*{arg\,min}_{\bm{w}}
\left|U^{\mathrm{learn}}\bm{w}-\bm{T}^{\mathrm{learn}}\right|^2,
\end{equation}
which is computed using the Moore--Penrose pseudoinverse.  The trained output for the test sequence is then
\begin{equation}
\bm{p}=U^{\mathrm{test}}\bm{w}.
\end{equation}
The capacity for a target sequence is evaluated from the squared correlation coefficient between the target $\bm{T}$ and the trained output $\bm{p}$,
\begin{equation}
C[\bm{T}]
=
\frac{
\left[\sum_{i=1}^{N}(T_i-\avg{T})(p_i-\avg{p})\right]^2
}{
\left[\sum_{i=1}^{N}(T_i-\avg{T})^2\right]
\left[\sum_{i=1}^{N}(p_i-\avg{p})^2\right]
}.
\label{eq:app_capacity}
\end{equation}

For the STMC, the target sequence is the $d$-step delayed input,
\begin{equation}
T_i^{(d)}=q_{i-d}.
\end{equation}
The memory capacity for delay $d$ is $C_d=C[\bm{T}^{(d)}]$, and the total STMC is defined as
\begin{equation}
{\mathrm{STMC}}
=
\sum_{d=1}^{d_{\max}} C_d.
\label{eq:app_stmc}
\end{equation}
In this work, we used $N=1000$ and $N_{\mathrm{node}}=50$ for the STMC calculation.  The input sequence was generated from a binary random function taking the values 0 or 1.

For the IPC, the target functions are nonlinear polynomial functions of past inputs.  Following the standard IPC framework, we construct orthogonal target sequences using Legendre polynomials $P_k(\cdot)$.  For a multi-index $\bm{a}=(a_1,a_2,\ldots,a_{d_{\max}})$, the target sequence is
\begin{equation}
T_i^{(\bm{a})}
=
\prod_{d=1}^{d_{\max}} P_{a_d}(q_{i-d}).
\label{eq:app_ipc_target}
\end{equation}
The set of target functions is restricted to
\begin{equation}
S(d_{\max},K_{\max})
=
\left\{
\bm{a}\in\mathbb{Z}_{\ge0}^{d_{\max}}
\;\middle|\;
0<\sum_{d=1}^{d_{\max}}a_d\le K_{\max}
\right\}.
\end{equation}
The capacity of each nonlinear target is evaluated using Eq.~(\ref{eq:app_capacity}), and the IPC is defined as the sum over the target set,
\begin{equation}
\mathrm{IPC}
=
\sum_{\bm{a}\in S(d_{\max},K_{\max})}
C[\bm{T}^{(\bm{a})}].
\label{eq:app_ipc}
\end{equation}
In the present IPC calculation, we used $N=5000$, $N_{\mathrm{node}}=50$, $d_{\max}=4$, and $K_{\max}=3$.  The input sequence was generated from a uniform random distribution over $[-1,1]$, for which the Legendre-polynomial targets form an orthogonal basis.  The values of $d_{\max}$ and $K_{\max}$ were chosen such that the number of target functions does not exceed the number of virtual nodes while still covering low-order nonlinear memory effects relevant to the present reservoir.
These parameters were selected as representative values for comparing the conventional VSTO and the m-VSTO under the same protocol, and they were not obtained by a systematic global optimization over all reservoir hyperparameters.

\FloatBarrier

\FloatBarrier

\end{document}